\documentstyle[epsfig,aps,floats,preprint,tighten,oldlfont]{revtex}

\def\beq{\begin{equation}}
\def\eeq{\end{equation}}
\def\Rvec{{\bf R}}
\def\rvec{{\bf r}}
\def\kvec{{\bf k}}
\def\half{{\textstyle{1\over 2}}}

\def\etal{{\it et al}}
\def\phibol{\mbox{\boldmath$\phi$}}
\def\sigbol{\mbox{\boldmath$\sigma$}}
\def\taubol{\mbox{\boldmath$\tau$}}
\def\shalf{{\scriptstyle{1\over 2}}}

\begin{document}
\draft
\preprint{MC/TH 97/16\kern 1 cm nucl-th/9710021}
\title{A self-consistent approach to the Wigner-Seitz treatment\\
of soliton matter}
\author{Urban Weber\footnote{Current address: Institute of Experimental
Physics, FR 10.2 Experimental Physics, Universit\"at des Saarlandes, D66041
Saarbr\"ucken} 
and Judith A. McGovern\footnote{Electronic address: j.mcgovern@man.ac.uk} }
\vskip 20pt
\address{Theoretical Physics Group, Department of Physics and Astronomy\\
University of Manchester, Manchester, M13 9PL, U.K.}
\nopagebreak
\maketitle

\begin{abstract}
We propose a self-consistant approach to the treatment of nuclear matter as a
crystal of solitons in the Wigner-Seitz approximation.  Specifically, we use a
Bloch-like boundary condition on the quarks at the edge of a spherical cell
which allows the dispersion relation for a given radius to be calculated 
self-consistently along with the meson fields; in previous work some ansatz
for the dispersion relation has always been an input.  Results in all models
are very sensitive to the form of the dispersion relation, so our approach
represents a significant advance.                       

We apply the method to both the Friedberg Lee model and the chiral quark-meson
model of Birse and Banerjee.  Only the latter shows short range repulsion; in
the former the transition to a quark plasma occurs at unrealistically low
densities.
\end{abstract}

\section{Introduction}

It is universally accepted that nucleons are composite objects made out of 
quarks, and that it will eventually be necessary to include this substructure 
in order to successfully model all properties of nuclei and nuclear matter.
For many years now there has been a program of modelling the nucleon as
being primarily composed of three light relativistic quarks bound together in
a localised spherical region by a mean field generated by their mutual
interaction.  Such soliton models have the advantage over their precursor,
the MIT bag, that the structure is fully dynamical, and so its response to
external probes can be calculated consistently.

Apart from the calculation of the properties of static, isolated nucleons, one
of the earliest applications of soliton models was to nuclear matter. There are
two main ways in which this has been done.  One, which was started later but
has been taken further, involves modelling the effect of the presence of other
nucleons on a test nucleon by constant background fields, typically the
$\sigma$ and $\omega$ mesons which play such an important role in  quantum
hadrodynamics \cite{Tjon97}.  Since these background fields will modify the
soliton structure, which in turn will change the meson-nucleon coupling
constants, the value of the fields for a given density can be calculated 
self-consistently.  Such calculations may well capture the essential physics at
low densities. At nuclear matter densities, however, the inter-nucleon
separation is very comparable to the nucleon radius.  If soliton models, in
which the quark profiles fall to zero smoothly as a function of distance from
the centre, are to be taken seriously, then the quark wavefunctions of
different nucleons  must overlap significantly.  The other approach to soliton
matter treats this as the dominant effect
\cite{achtzehnter85,birse88,banerjee85,glen86,dodd87,fai96}.               

If nuclear matter were crystalline, we would know how to tackle the problem.
Single quark energy levels would be replaced by Bloch wavefunctions, satisfying
$\psi(\rvec+\Rvec)=e^{i\kvec\cdot\Rvec}\psi(\rvec)$ for lattice vectors 
$\Rvec$.  By concentrating on a single cell the Bloch condition gives boundary
conditions on the wavefunction within that cell for all values of the crystal
momentum $\kvec$ within the first Brillouin zone, with the lowest energy state
corresponding to $\kvec=0$ and an energy gap above the highest energy state. 
It is therefore possible to start with an approximate form of the potential,
find a sufficient set of states to enable the quark density to be
integrated over the band, solve for the potential generated by these quarks 
and then iterate until convergence is reached.  While simple in principle,
this would be a formidable task in practice, since the potential and
wavefunctions will contain all spherical harmonics and the integration in
$\kvec$-space is  three-dimensional.  The closest to realising this program
which has been  achieved so far was also the earliest such study. Achtzehnter 
\etal\ \cite{achtzehnter85} used the Friedberg-Lee model on a face-centred
cubic lattice.  However they calculated states and energies for only one
direction in $\kvec$-space, and assumed that the dispersion relation was the
same in all directions, thereby simplifying the calculation of the quark
density.
                                         
However nuclear matter has no long-range crystalline order.  This fact, 
coupled with a desire for a less cumbersome calculational method, inspired the
approach that has been taken ever since, namely the use of the Wigner-Seitz
approximation \cite{wigner33}.  In their first paper Wigner and Seitz were
concerned with the energy of the bottom of the band in sodium. They made the
assumption that within the cell the potential is spherically symmetric,  and
also that the polyhedral cell boundary is well approximated by a sphere. The
Schr\"odinger wavefunction for the state at the bottom of the band will 
therefore be spherically symmetric and be flat on the boundary. (For Dirac
wavefunctions, those become the conditions on the upper component.)   These
assumptions seem particularly suitable for soliton matter, which is not
crystalline, since they do not depend on any particular lattice, but instead
reflect the average isotropy of a liquid.  For a full calculation, however, it
is not enough to have the bottom of the band; the wavefunctions of all states
in the band are required.  

Most authors in the field of soliton matter have
assumed that all the states in the band are also are $s$-wave states
(surprisingly, since the Bloch condition is strongly anisotropic for  non-zero
$\kvec$) so that the dispersion relation is independent of direction.  They
then made a variety of assumptions about the form of the dispersion relation
and the position of the top of the band. Birse \etal\ \cite{birse88} for
instance took the top to be the state for which the  upper component of the
Dirac wavefunction vanishes at the cell boundary.  Between these two states
the dispersion relation was assumed to be a sinusoid, with $d \epsilon/d k=0$
at the bottom and top of the band.  The quark density was built up by
sampling the full band.  Another assumption used has been that the width of
the band is twice the difference between the energy of the isolated soliton
and the energy of the bottom of the band defined as  above \cite{banerjee85}.
A third is to assume that the dispersion relation is given by 
$\epsilon=\sqrt{\epsilon_b^2+k^2}$, with the top of the band given by 
$k_t=\pi/2R$ \cite{glen86}.  The last assumption is also used in 
ref.~\cite{fai96}, where, following Wigner and Seitz's second paper, they
write $\psi_{\bf k}(\rvec)  =e^{i\kvec\cdot\rvec}u_{\bf k}(\rvec)$, and then
assume that $u_{\bf k}(\rvec)= \psi_{\bf 0}(\rvec)$.  Wigner and Seitz
carefully checked this assumption---which is clearly not exact since the
equation for $u_{\bf k}(\rvec)$ depends on $\kvec$---and found that
for sodium, where the quadratic dispersion relation reflects the fact that the
electrons are nearly free, the approximation was a good one.  For
well-separated solitons, in which the quarks remain clearly localised  and the
dispersion relation is likely to show pronounced flattening at the top of the
band, it is unlikely to hold. Some authors have simplified still further by
allowing only the states at the  top or bottom of the band to be occupied. The
problem with all these approaches is that the algorithm for finding the top of
the band is completely ad-hoc. Furthermore the contribution of the quarks to
the energy of the cell, proportional to $\int_0^{k_t} \epsilon(k)\,k^2dk$, is
dominated by the higher states, and so the results for the dependence of the
energy on density are extremely sensitive to the assumptions used.  None of
the studies performed so far  can thus be regarded as reliable guides to
physics.    
                                           
The current work was motivated by the desire to retain the strengths of the
Wigner-Seitz approach, namely the lack of dependence on a particular lattice
structure and the recognition of the approximately spherically-symmetric
environment of the nucleon, but in a fashion that allows the band structure to
be calculated self-consistently.  We therefore propose to retain a Bloch-like
boundary condition to relate the values of the wavefunction at any pair of 
antipodal points for points on the boundary $\cal S$,
\beq
\psi(\rvec)=e^{2i\kvec\cdot\rvec}\psi(-\rvec), \qquad\forall \rvec\in {\cal S}.
\label{bc}
\eeq
The closer to spherical the Brillouin zone for a particular lattice is, the 
closer this is to the normal Bloch condition. Essentially this is the logical
continuation of the Wigner-Seitz method.  Furthermore this boundary condition
is also satisfied by a plane wave of momentum $\kvec$, and so it  also holds
for the quark plasma state which will be the lowest energy state at high
densities.   For a band based on an $s$-wave state, this gives the same
condition for the bottom of the band, $k=0$, as was used previously, namely
the vanishing of the lower component at the cell boundary  (being a $p$-wave,
the lower component is odd under $\rvec\to-\rvec$).  However for a given
spherical potential this condition can be solved for any $\kvec$. If the
potential is sufficiently deep, this  gives a dispersion relation
$\epsilon(k)$ which turns over  at a  certain value of $k$; this point, at
which $d\epsilon/dk=0$, is naturally taken as the top of the band. (The
dispersion relation is of course isotropic.)   The band based on the next
state (with a radial node in the isolated wavefunction) will not overlap with
the lowest band, and so there is a band gap.   The wavefunctions will not of
course be spherically symmetric, except at the bottom of the band, and so are
taken as  sums over partial waves. However the quark density, integrated over
the direction as well as the magnitude of $\kvec$, {\it is} spherically
symmetric, and so can be used to generate a new spherically symmetric
potential; the process is repeated until convergence is reached.  This is the
first fully self-consistent calculation of soliton matter to be reported.
                                         
There are physical effects which have not been taken into account in this
calculation; spurious centre of mass energy has not been subtracted nor energy
due to fermi motion of the nucleons added, and the effects of one-gluon
exchange or meson clouds apart from those which form the soliton have not been
included---the effects which split the nucleon from the delta are absent.
This is not yet a realistic model of nuclear matter, but it is a better
starting point than any proposed so far. 

\section{The Models}

Two models have been used in the current work: the Friedberg-Lee model and 
the chiral quark-meson model, which we will introduce briefly in turn.

The Friedberg-Lee model \cite{FL77} is the prototypical soliton model,
consisting only of quarks and a single scalar field, $\sigma$, which couples
linearly to the quarks.   The potential for the $\sigma$ has an absolute
minimum at  a non-zero value of $\sigma$, $\sigma_v$, which is the value the
field takes in the vacuum, but in the presence of a scalar quark density the
minimum of the effective potential is pushed closer to zero.  While quarks can
exist as free particles, with a mass $g\sigma_v$, it will be energetically
favourable for three quarks to be localised in space in a ``bag" of lower
$\sigma$; this is a soliton.  The Lagrangian is as follows:
\beq
{\cal L}={\overline\psi}\left(i\gamma^\mu \partial_\mu
-g\sigma\right)\psi +\half\partial_\mu \sigma \partial^\mu \sigma -U(\sigma),
\eeq
where $U(\sigma)=a\sigma^2/2+b\sigma^2/3!+c\sigma^4/4!-B_{vac} $, 
and results will be 
presented for the case $a=0$, $b=-700.43$~fm$^{-1}$, $c=10^4$ and $g=10.98$,
giving $B_{vac}=0.27$~fm$^{-4}$ and $\sigma_v=0.21$~fm$^{-1}$.  For these
parameters the isolated soliton has an isoscalar r.m.s.\ radius of 0.83~fm,
and energy of 1265~MeV.

The Friedberg-Lee model lacks chiral symmetry.  This can be restored by making
the $\sigma$ field the chiral partner of the pion potential. This chiral 
quark-meson model was introduced by Birse and
Banerjee \cite{BB84} and by Kahana \etal\ \cite{KRS84}. The Lagrangian is 
\beq
{\cal L}={\overline\psi}\left(i\gamma^\mu \partial_\mu
+g(\sigma+i\gamma_5\taubol\cdot\phibol)\right)\psi 
+\half\partial_\mu \sigma \partial^\mu \sigma 
+\half\partial_\mu \phibol \partial^\mu \phibol -U(\sigma,\phibol),
\eeq
where $U(\sigma,\phibol)$ is the Mexican hat (wine-bottle) potential, quadratic
in $\sigma^2+|\phibol|^2$, plus a small symmetry breaking term linear in
$\sigma$ which gives explicit chiral symmetry breaking and a non-zero pion mass.
Unlike the F-L model, for which soliton formation is independent of the
spin-isospin structure of the quarks, which may therefore be taken to be in 
the appropriate SU(6) wavefunction, spherical solutions in the chiral model
require all three quarks to be in a ``hedgehog" configuration,
$\chi_h=(u\downarrow-d\uparrow)/\sqrt{2}$.  The $\sigma$ field is radially
symmetric, and the pion fields have their isospin vector pointing radially
outwards, $\phi_i(\rvec)=\hat{r}_i\,h(r)$.  The resulting soliton is a
superposition of all states with equal spin and isospin, namely the nucleon, 
delta and higher resonances.  States of good spin and isospin must be
projected out, by cranking or other methods.  It is far from clear how to do
this for a soliton in matter, and we here deal only with hedgehog matter in
this model.  We use parameters in the potential which give 
$\sigma_v=f_\pi=93$~MeV, $m_\pi=139$~MeV and $m_\sigma=1200$~MeV, and take 
$g=5.38$, which gives a soliton energy of $1116$~MeV. 

\section{Soliton matter: the method}

For each cell radius, $R$, a self-consistent solution was obtained, 
consisting of a
spherically symmetric $\sigma$ field subject to the boundary condition
$\sigma'(R)=0$ and quarks in states characterised by the crystal momentum
$\kvec$, satisfying the boundary condition (\ref{bc}), with states evenly
distributed throughout a sphere in $\kvec$-space of radius $k_t$, $k_t$ being
the top of the band.  As stressed in the introduction, the dispersion relation
$\epsilon(k)$  and $k_t$ are themselves calculated self-consistently.  If the
pion field is  present, its odd parity leads to the boundary condition $h(R)=0$.

Details of the calculation are as follows.  We start with an isolated soliton,
in order to have a good approximation to the potential at large radii, and
after a converged solution is found for each radius its potential is used
as the initial guess for a smaller radius; thus we work inwards until the
solution is finally lost.

The quark scalar density is found as follows for the F-L model. The hedgehog 
ansatz introduces some complications which are covered in the appendix. For a
given potential, the bottom of the band is easily found by solving the Dirac
equation for the $s$-wave state subject to the condition that the lower
component vanishes at $r=R$.  In order to have what turns out from experience
to be an upper limit on the energy of the top of the band, it is useful also
to find the energy of the state for which the upper component vanishes at $R$. 
The true  top is typically about half way between the bottom and this upper
limit. The wavefunction for non-zero $\kvec$ will not be spherically
symmetric, but a superposition of states of all angular momenta; it proves
sufficient in practice to include the first 14 states.  Since the boundary
condition is independent of spin and axially symmetric, states with spin
projection parallel and antiparallel to $\kvec$ have the same energy; we need
only consider one of them and allow for spin and isospin degeneracy
separately.  Thus, taking $\kvec=k\hat{\bf z}$, we use for  our trial state  
\beq
\psi_k(\rvec)=\sum_\kappa i^{l_\kappa} D_\kappa 
\left(u_{\kappa,\epsilon}(r)\atop
i\sigbol\cdot\hat{\rvec}\, v_{\kappa,\epsilon}(r) \right) 
\Phi^\shalf_\kappa(\theta,\phi).
\label{wavefn}
\eeq
The quantum number $\kappa$ ($\kappa=\pm 1,\pm 2,\ldots$) gives the total
angular momentum, $j_\kappa=|\kappa|-\half$ and the orbital angular momentum of
the upper component, $l_\kappa=\kappa$ for $\kappa>0$ and $l_\kappa=|\kappa|-1$
for $\kappa<0$; $\kappa=-1$ corresponds to the $s$-wave state. The
$\Phi^m_\kappa(\theta,\phi)$ are spin-orbit functions of spin-$\half$ and 
orbital angular momentum $\l_\kappa$ coupled up to $j_\kappa$: 
\beq
\Phi^m_\kappa=\sum_{m_l,m_s}
\langle l_\kappa\half m_lm_s|j_\kappa m\rangle Y_l^{m_l}\chi_{m_s},
\eeq
and satisfy the relation
$\sigbol\cdot\hat{\rvec}\,\Phi^m_\kappa=\Phi^m_{-\kappa}$.
The sum in (\ref{wavefn}) is over all the eigenstates of angular momentum of a
given energy $\epsilon$ in the potential ($\epsilon$ can be chosen at will
because the only boundary condition imposed on each partial wave is that the
wavefunction is regular at the  origin.) The power of $i$ is extracted to make
the coefficients $D_\kappa$ real.  The radial functions
$u_{\kappa,\epsilon}(r)$ and $v_{\kappa,\epsilon} (r)$ satisfy the Dirac
equations            
\begin{eqnarray}
&&u'+{\kappa+1\over r}u+(\epsilon+g\sigma)v=0,\nonumber\\
&& v'-{\kappa-1\over r}v-(\epsilon-g\sigma)u=0.
\end{eqnarray}

The wavefunction (\ref{wavefn}), with an arbitrarily chosen energy, will not
satisfy the boundary condition (\ref{bc}) for an arbitrary $k$.  However
for any $k$ there will be an energy for which the boundary condition is
satisfied, and conversely for a given energy, if it is not in a band gap, there
will be  a corresponding value of $k$.  Finding the correspondence determines
the  dispersion relation $\epsilon(k)$. This was done as follows.  The boundary
condition (\ref{bc}) can be rewritten
\beq
e^{-ikR \cos\theta}\psi(R,\theta,\phi)-e^{ikR \cos\theta}\psi(R,\pi-\theta,
\pi+\phi)=0.
\eeq
The exponentials can be expanded in terms of Legendre polynomials and spherical
Bessel functions, and we substitute in the form (\ref{wavefn}) for $\psi$. If
we  multiply by the spherical harmonic $Y^{M}_L$, for any $L$ and $M=0$ or 
$-1$,  and integrate over solid angle, we obtain an equation for a linear
combination of the coefficients $D_\kappa$.  The set of these can be written as 
a matrix equation as follows, where the rows of the matrix are labelled by
$(L,M)$:
\beq
\sum_\kappa {\bf M}_{{\scriptscriptstyle(L,M)}\kappa}D_\kappa=0,
\eeq
where
\beq
{\bf M}_{{\scriptscriptstyle(L,M)}\kappa}(k,\epsilon)=
{\bf m}_{{\scriptscriptstyle(L,M)}\kappa}(k)
\left\{ {v_{\kappa,\epsilon}(R)  \atop u_{\kappa,\epsilon}(R)}\right\} 
\hbox{for} \left\{ \hbox{$L$ even} \atop   \hbox{$L$ odd}\right.
\eeq
and
\begin{eqnarray}
&&{\bf m}_{{\scriptscriptstyle(L,M)}\kappa}(k)=\sum_l j_l(kR)(2l+1)
|\kappa-M|^\half \left({\rm Sign}(\kappa)\right)^M (-1)^{(l+l_\kappa+L)/2}
\nonumber \\
&&\kern 2cm \times\left(\matrix{l&L&l_\kappa\cr 0& 0 &0}\right)
 \left(\matrix{l&L&l_\kappa\cr 0& M &-M}\right).
\label{matrix}
\end{eqnarray}

Numerically, if we truncate the expansion of the
wavefunction at $2n$ terms, we use only the first $n$ values of $L$, giving
a $2n\times 2n$ matrix.
This matrix equation can hold for non-trivial $D_\kappa$ only if
det$\bigl(M(k,\epsilon)\bigr)= 0$. In practice we search for 
the value of $k$ which gives a vanishing determinant for a given $\epsilon$. 
The top of the band, $k_t$, corresponds to the highest value of $\epsilon$ for
which such a $k$ can be found.  A typical dispersion relation is shown in
figure 1.

\begin{figure}
  \begin{center}
\mbox{\kern-1.0cm
\epsfig{file=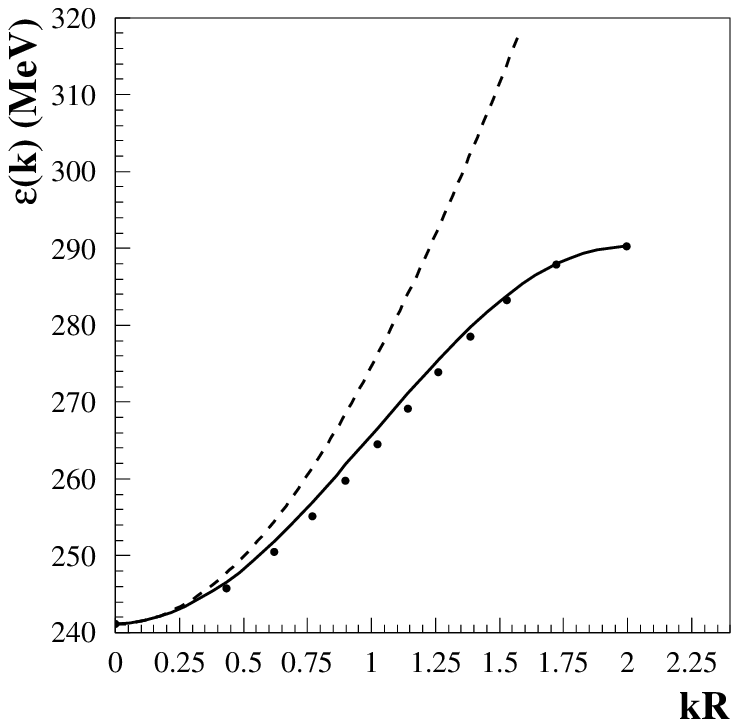,width=9.0truecm,angle=0}}
  \end{center}
\vspace{-1.5 cm}

{\bf Fig.~1:} Points: the dispersion relation for the F-L model with cell radius
1.5~fm. Solid curve: a sinusoidal fit to the data points.  Dashed curve:
the dispersion relation $\epsilon(k)=\sqrt{e_b^2+k^2}$ as assumed in 
Refs~\cite{banerjee85,fai96}.
\end{figure}

Having found the dispersion relation for $\kvec$ along the $z$-axis, we know
from spherical symmetry that it will be the same in any other direction. A
simple rotation serves to produce the corresponding wavefunctions.  The full
scalar density is then spherically symmetric, and is given in terms of the
$\kvec=k\hat{\bf z}$ solutions by                         
\beq
\rho_s(r)=(3/k_t^3)\int_0^{k_t}k^2\sum_\kappa D_\kappa^2
\left(u_{\kappa,\epsilon(k)}(r)^2-v_{\kappa,\epsilon(k)}(r)^2\right)\,dk.
\label{scdens}
\eeq
The integration is done using Simpson's rule, with values of the integrand  at
regular values of  $k$ found by interpolation from the set of solutions,
generally regularly spaced in $\epsilon$, which was used to find the dispersion
relation. An overall normalisation constant has been included in the
$D_\kappa$'s  so that the corresponding vector density, integrated over the
sphere, is three.

The scalar density then enters as a source term in the mean-field equation 
for the $\sigma$ field.  With a new field, the steps can be repeated until
convergence is obtained.  The total energy is then
\beq
E=(9/k_t^3)\int_0^{k_t}\epsilon(k)k^2 dk+4\pi\int_0^R
r^2\left(\half\sigma'(r)^2+U(\sigma)\right)dr.
\label{flenergy}  
\eeq

In everything above we have assumed that the band is filled right to the top.
This is not in fact obvious.  Isospin, spin and colour give each state a
twelve-fold degeneracy, and there are only three quarks per soliton; the
minimum energy state would therefore correspond to filling only the bottom
quarter of the states.  However such a filling would in no way correspond to
the quarks in any one cell being coupled up to a colour singlet with the
quantum numbers of the proton.  Furthermore gluon exchange energies are
comparable with the width of the band, and so are capable of negating any
saving in energy produced by a concentrated filling.  Following
ref.~\cite{birse88} we have used dilute filling.  

In the case of the chiral quark-meson model the quark states are characterised
by grand spin (sum of spin and isospin), and the filled quark band gives rise
not only to a spherically symmetric quark density, but also an isovector 
pseudoscalar density.  As required for self-consistency, this has indeed the 
hedgehog form $\hat{r}_i\,\rho_h(r)$.  However in practical calculations
the full scalar density in the F-L model was found to be very well approximated
by truncating the sum over $\kappa$ in (\ref{scdens}) at the lowest value,
$\kappa=-1$, and so in the chiral model we only used the $G=0$ states in
calculating the source terms for the meson fields. The quark energy has the
same form as Eq.~(\ref{flenergy}), and the meson energy is 
\beq                                                
E_{\sigma\phi k}+E_{\sigma\phi p}= 4\pi\int_0^R
r^2\left(\half\sigma'(r)^2+\half h'(r)^2+U(\sigma,h)\right)dr.
\label{mesenergy}
\eeq
The derivative and potential terms have been labeled separately for future
reference. With hedgehog quarks the only degeneracy is due to colour, and so
there is no ambiguity in the filling  procedure; the band is filled to the top.

\section{Results}

\subsection{Friedberg-Lee Model}

\begin{figure}
  \begin{center}
\mbox{\kern-1.3cm
\epsfig{file=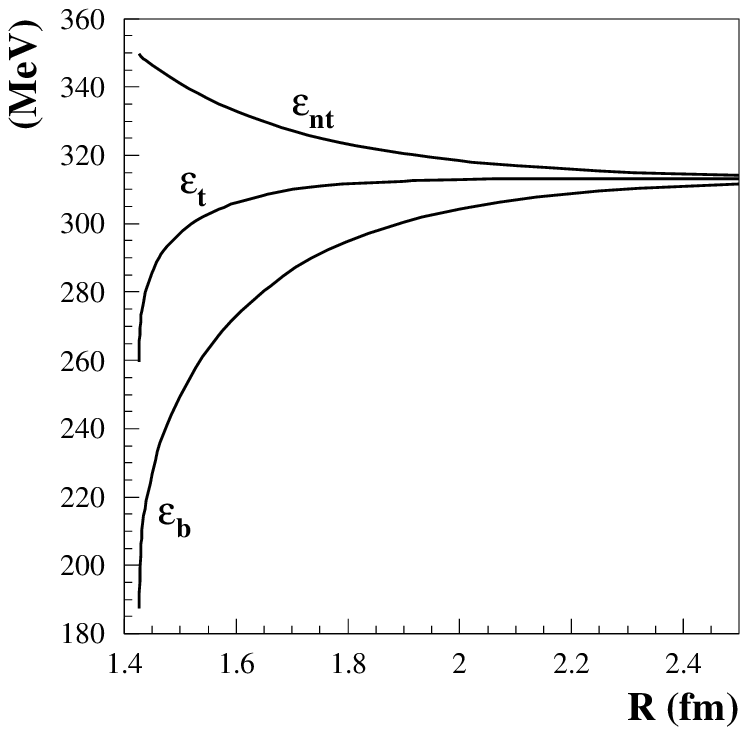,width=9.0truecm,angle=0}\kern-1cm
\epsfig{file=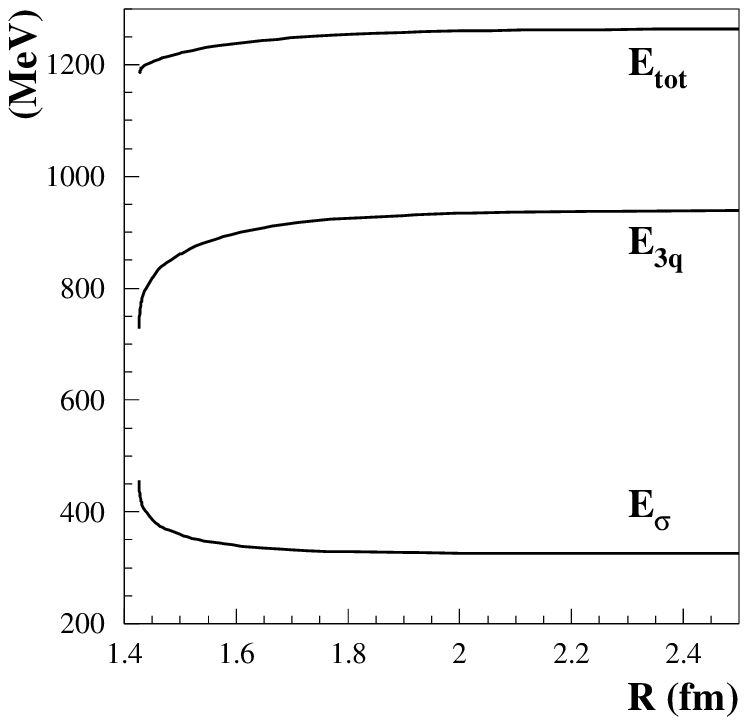,width=9.0truecm,angle=0}}  \end{center}
\vspace{-1.5 cm}

{\bf Fig.~2:} The top and bottom of the band ($\epsilon_t$ and $\epsilon_b$)
and the total energy $E_{tot}$ along with its decomposition into quark
($E_{3q}$) and mean field ($E_{\sigma}$) parts.  Also shown is the top of the
band estimated by the method of \cite{birse88} ($\epsilon_{nt}$), which is not
used in this calculation.
\end{figure}

Figure 2 shows the energy of a single cell as a function of cell radius. 
As  the radius becomes large, the band width shrinks to zero and the total
energy  tends to the value for an isolated soliton.  As the radius decreases,
the band  widens.  The bottom of the band falls fast, but the top falls too,
and as a result the quark energy falls quite rapidly.  The $\sigma$ energy
rises, but  the overall energy drops.
                          
\begin{figure}
  \begin{center}
\mbox{\kern-1.3cm
\epsfig{file=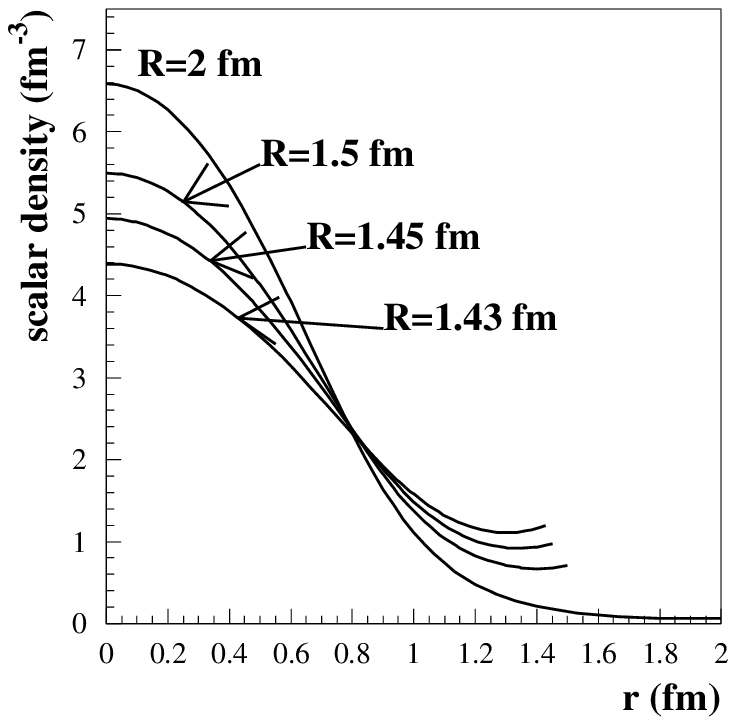,width=9.0truecm,angle=0}\kern-1cm
\epsfig{file=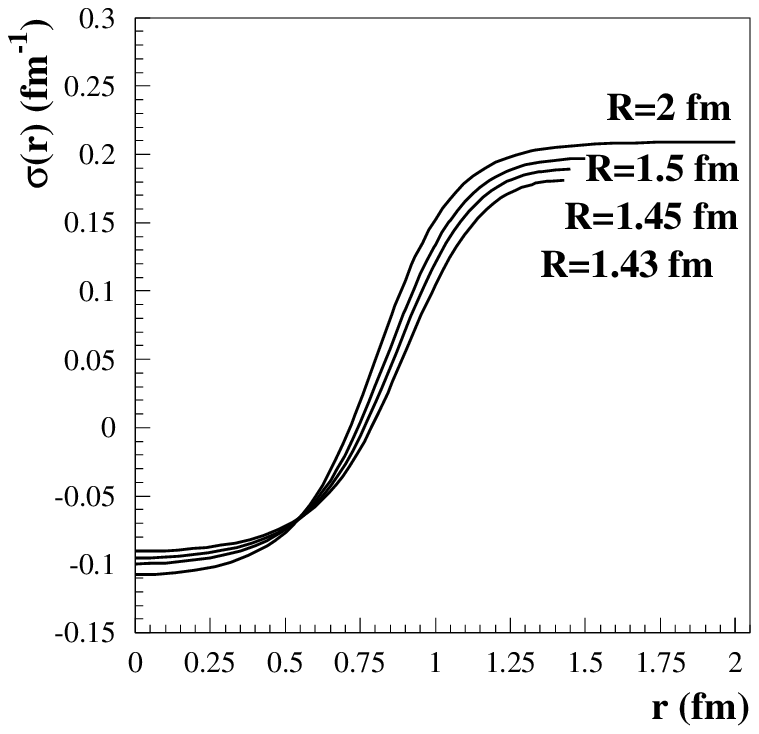,width=9.0truecm,angle=0}} \end{center}
\vspace{-1.5 cm}

{\bf Fig.~3:} The quark scalar density and meson field profiles for
different radii in the F-L model.
\end{figure}

Figure 3 helps to explain this: as the cell radius decreases, the mean
radius of the quarks actually increases, with a corresponding decrease in
kinetic energy which dominates the energy of the whole cell. We note also from
figs~1 and 2 that the estimate of the top of the band given by the energy at
which the  upper component vanishes at the cell boundary ($\epsilon_{nt}$), or by
$\sqrt{e_b^2+(\pi/2R)^2}$, is very different to that
given  by the self-consistent boundary conditions used here. This has a
profound effect on the results, which are quite different.
              
Dodd and Lowe \cite{dodd91} have studied the one-dimensional F-L soliton
crystal analytically.  They found that the energy of the lowest-energy soliton
state 
was very close to the
energy of the massive-quark plasma, the difference becoming very small as the 
cell radius decreases.  Unless the top state of the band was occupied, there
was a radius at which the solution actually bifurcated from the plasma
solution.  
In our calculations, however, the soliton solution is
well below the massive plasma branch.  
The solution is finally lost at a radius of 1.427~fm, at which point it is
clear from fig.~2 that the cell cannot be shrunk further. (At this point the
band gap is still wider than the band.) This radius, at
which the density is well below nuclear matter density, is disappointing.
However it is not surprising.  The Friedberg-Lee model has a massless plasma
phase which is unrealistically low in energy (at a density corresponding to a
radius of 1~fm, the massless plasma with a twelve-fold degeneracy of quark
states has an energy of only 900~MeV per three quarks.)  At the radius at which
the solution is lost, a plasma with the same value of $k_t$ is already of lower 
energy than our solution.

The calculation has also been done for the two parameter sets of
ref.~\cite{achtzehnter85},
which were tuned to give the  same r.m.s.\ quark radius for the isolated
soliton; the results were not notably different.

It is worth commenting on the wavefunctions obtained with this method. Clearly
it is of the essence that, for $\kvec\ne 0$, they are not pure $s$-waves,
though in fact the admixtures of higher states are small.  Nearly identical
results are obtained if these are ignored in calculating the scalar density.
However contrary to the assumption of ref.~\cite{fai96}, the contributions of
states with different $k$ to the scalar density are markedly different, and as
the high $k$ states dominate because of the $k^2$ factor, quite different
results would be obtained if only the $k=0$ state were used.
 
\subsection{Chiral quark meson model}

\begin{figure}
  \begin{center}
\mbox{\kern-1.3cm
\epsfig{file=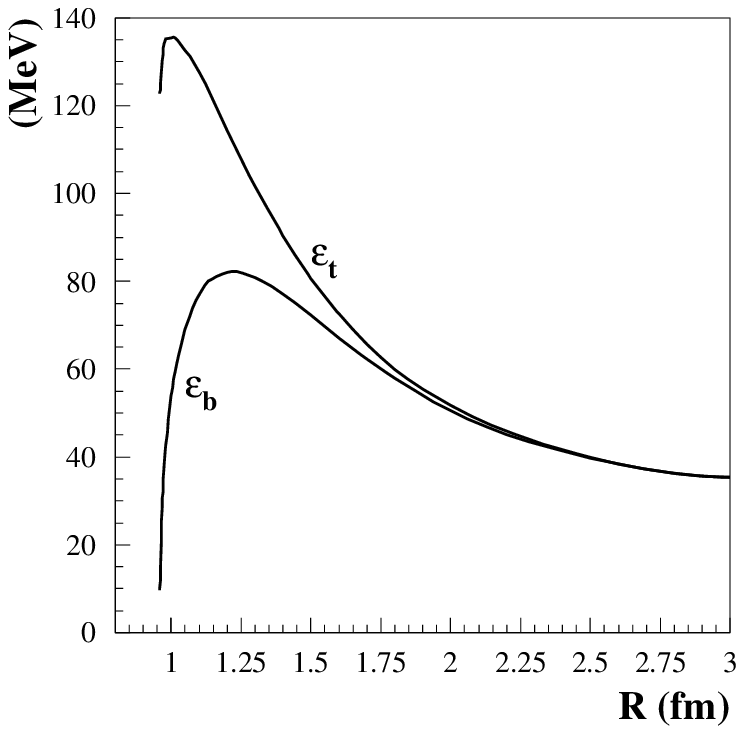,width=9.0truecm,angle=0}\kern-1cm
\epsfig{file=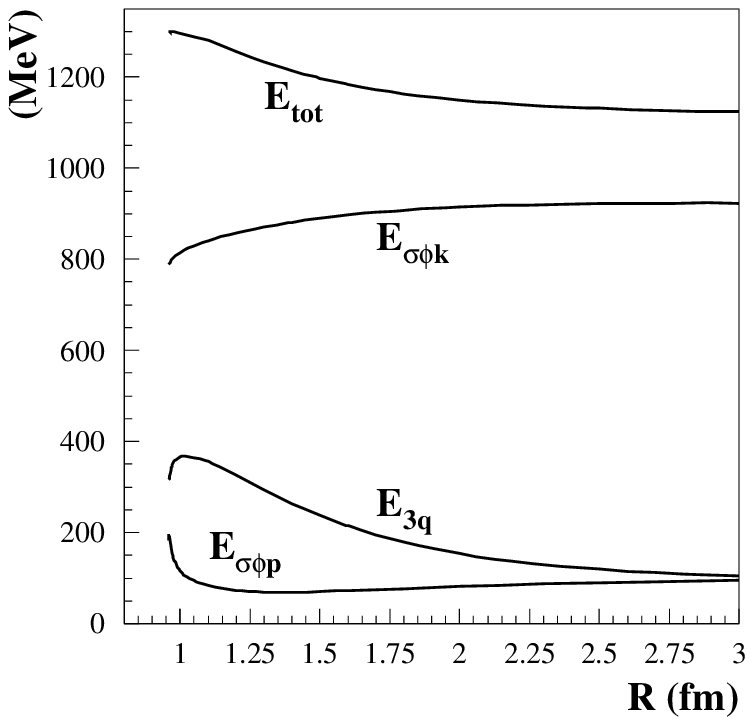,width=9.0truecm,angle=0}}  \end{center}
\vspace{-1.5 cm}

{\bf Fig.~4:} The top and bottom of the band ($\epsilon_t$ and
$\epsilon_b$) and the total energy ($E_{tot}$) along with its decomposition into
quark energy ($E_{3q}$) and derivative and potential meson-field energies
($E_{\sigma\phi k}$ and $E_{\sigma\phi p}$, Eq.~(\ref{mesenergy})), in the
chiral quark-meson model.  
\end{figure}

The behaviour of the chiral model (figs~4 and 5) is quite
different form that of the Friedberg-Lee model.  Until just below nuclear
densities ($R=1.07$~fm), both the bottom and top of the band rise with
decreasing radius, as does the total energy, in spite of the fact that the
mesonic energy tends to fall.  In contrast to the F-L model it can be seen that
the mean radius of the quarks  remains fairly constant, while the meson fields
start to deviate from the chiral circle, increasing the potential energy of the
quarks.  One might expect the  potential energy of the mesons ($E_{\sigma\phi
p}$) to rise also, but this is  clearly offset by the decreasing volume of the
cell. At nuclear densities  there is a dramatic change in behaviour, as the
quark energy falls sharply and the total energy starts to level off.  At 
$R=0.958$~fm the solution is lost.
                                                
\begin{figure}
  \begin{center}
\mbox{\kern-1.3cm
\epsfig{file=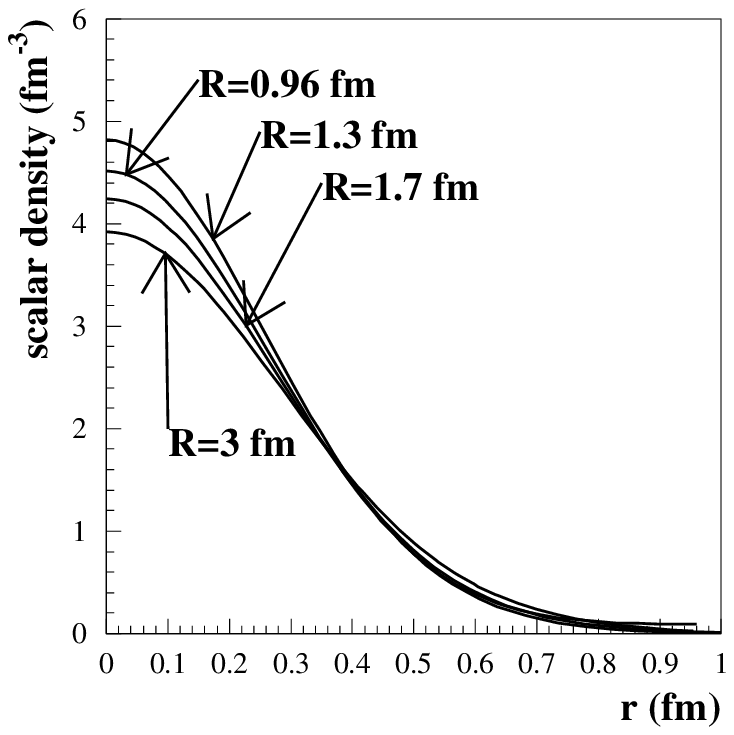,width=9.0truecm,angle=0}\kern-1cm
\epsfig{file=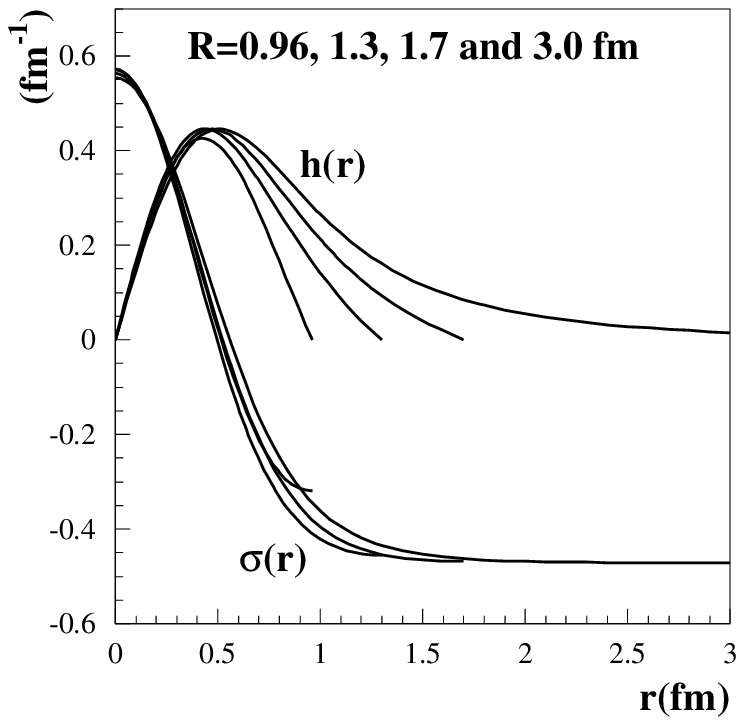,width=9.0truecm,angle=0}} \end{center}
\vspace{-1.5 cm}

{\bf Fig.~5:} The quark scalar density and meson field profiles for
different radii.
\end{figure}

This time, when the solution vanishes, the plasma solution (with a uniform 
$\sigma$ and zero pion field) is still much higher in energy, since the meson
field are sitting at the top of the Mexican hat rather than approximately
following the brim, as they do in the soliton.   There are lower plasma states
with plane-wave pion fields  \cite{Broniowski90}, but these do not satisfy
our boundary conditions.   However there may be another, plasma-like solution
in which the meson fields do not wrap round the chiral circle but make an
excursion round it before returning to their original values; the quarks  would
be in weakly-distorted plane-wave states.  Since there would be no band gap,
such a solution is cannot be found with our algorithm.  Transition to this
state would however represent deconfinement to a state of matter with a pion
condensate, which therefore happens at  1.4
times nuclear density in this model.  
             
In summary neither model shows the full behaviour expected of the
energy-density curve, which is expected to have a long-range attraction and
a short-range repulsion.  In view of the many physical effects omitted 
(meson exchange, gluonic effects, fermi-motion of nucleons) and  the
deficiencies of the models used, this is perhaps not surprising.

\section{Conclusion}

We have introduced a novel method of treating soliton matter in the
Wigner-Seitz approximation which avoids the arbitrariness of previous
treatments.  By imposing a Bloch-like boundary condition---which however is
independent of any crystal lattice---on the quarks in a spherically symmetric
potential we can generate the dispersion relation complete with band gaps, 
and by filling a band we provide a spherically symmetric source for the field
which gives rise to the potential.  Thus for the first time a completely 
self-consistent treatment of the fields in the Wigner-Seitz cell is possible.
The same boundary conditions are also 
satisfied by the uniform plasma solution which is the lowest energy state at
high densities, and so this method is particularly suited to examining the 
deconfinement transition.

Many previous Wigner-Seitz calculations of soliton matter have been carried
out.  All have found that the results are extremely sensitive to the choice
of the top of the band.  It is therefore very satisfactory to have a way of
determining this self-consistently.  

We have carried out calculations in the Friedberg-Lee and chiral quark-meson
models.  In the former no signs of repulsion are seen, and the solution
disappears well below nuclear densities as the plasma solution takes over as
the lowest energy state.   In the latter short-range repulsion is seen, 
and the solution is lost at around 1.4
times nuclear density.  Since the solitons in this case are hedgehogs, however,
rather than nucleons, the relevance to nuclear matter remains to be elucidated.

\acknowledgements

J. McG.\ would like to thank M. Birse for useful conversations and suggestions.
This work was supported in part by the UK EPSRC.  U. W. was supported by the
Studienstiftung des deutschen Volkes.

\appendix
\section*{The dispersion relation for hedgehog quarks}

The quantum number which commutes with the Dirac Hamiltonian in the chiral 
quark-meson model is grand spin, which is the sum of spin and isospin: $\bf 
G=I+j$.  Eigenstates of grand spin have been constructed by Kahana and Ripka
\cite{KR84}.  In the current case the projection of $\bf G$ along $\kvec$ is a
good quantum number, and since we are interested in the band which develops
from the isolated $G=0$ state, we need only consider $M_G=0$.  For a given $G$
there are four distinct Dirac spinors, two of each parity, and the coupling to
the mean fields mixes the same-parity pairs.  Thus we can write down
wavefunctions characterised by grand spin and parity: writing $q_G$ for states
with $P=(-1)^G$ and $p_G$ for states with $P=(-1)^{G+1}$, we have 
\begin{eqnarray}
q_G(\rvec)=\sqrt{{1\over2}}
\left[\left({u_1(r)\atop i\sigbol\cdot\hat{\rvec}\, v_1(r)} \right) 
\left(\Phi^{-\shalf}_{-G-1}u-\Phi^{\shalf}_{-G-1}d\right)+
\left({f_1(r)\atop i\sigbol\cdot\hat{\rvec}\, g_1(r)} \right) 
\left(\Phi^{-\shalf}_{G}u+\Phi^{\shalf}_{G}d\right)\right]
\end{eqnarray}                       
\begin{eqnarray}
p_G(\rvec)=\sqrt{{1\over2}}
\left[\left({v_2(r)\atop i\sigbol\cdot\hat{\rvec}\, u_2(r)} \right) 
\left(\Phi^{-\shalf}_{G+1}u-\Phi^{\shalf}_{G+1}d\right)+
\left({g_2(r)\atop i\sigbol\cdot\hat{\rvec}\, f_2(r)} \right) 
\left(\Phi^{-\shalf}_{-G}u+\Phi^{\shalf}_{-G}d\right)\right]
\end{eqnarray}                       
and the radial function obey the following equations

\def\us{u_{\scriptscriptstyle {1\atop 2}}}
\def\vs{v_{\scriptscriptstyle {1\atop 2}}}
\def\fs{f_{\scriptscriptstyle {1\atop 2}}}
\def\gs{g_{\scriptscriptstyle {1\atop 2}}}
\begin{eqnarray}
&&
\vs'+{G+2\over r}\vs-(g\sigma\pm\epsilon)\us\pm(-\vs\cos\theta+\gs\sin\theta)gh
=0 \nonumber\\&&
\us'-{G\over r}\us-(g\sigma\mp\epsilon)\vs\mp(-\us\cos\theta+\fs\sin\theta)gh=0
\nonumber\\&&
\gs'-{G-1\over r}\gs-(g\sigma\pm\epsilon)\fs\pm(\gs\cos\theta+\vs\sin\theta)gh=0
\nonumber\\&&
\fs'+{G+1\over r}\fs-(g\sigma\mp\epsilon)\gs\mp(\fs\cos\theta+\us\sin\theta)gh=0
\end{eqnarray}
where $\sin\theta=2\sqrt{G(G+1)}/(2G+1)$ and $\cos\theta=1/(2G+1)$.
(In deriving these equations, one should note that the convention of
ref.~\cite{KR84}
for the spin-angle functions is different from ours, with the result that
each element of their table for the matrix elements of $\taubol\cdot
\hat{\rvec}$ (Eq.~A.5) should be multiplied by $-1$ for use with our
wavefunctions.) For the ground state, $G=0$, there is only one,
positive parity, state, $q_0$, and only $u_1$ and $v_1$ are non-zero. Now
because each of the other wavefunctions is a sum of two spinors, there will be
two linearly independent solutions of the same form, which we obtain by
imposing two different boundary conditions at the origin; we choose to set each
of the two lower components to zero in turn to get the two different solutions. 
This extra degree of freedom requires an extra index, so that we have, for
$G\ne0$, $q_G^n$, $p_G^n$, where $n=1,2$. In constructing a wavefunction to
satisfy the boundary conditions (\ref{bc}) we use a superposition of states of
all grand spin and parity:
\beq
\psi(\rvec,t)=\sum_G\sum_{n=1}^2 i^G\left(B_G^n q_G^n(\rvec)+
C_G^n p_G^n(\rvec)\right)e^{-i\epsilon t}.
\label{hhwf}
\eeq
The boundary condition on the wavefunction is the same as before.
When imposed on  (\ref{hhwf}) the states with parity $(-1)^G$
and those with $(-1)^{G+1}$ decouple; since we want to consider the band based
on the ground state $G^P=0^+$, this means that we only need to include the  
$(-1)^G$ states in (\ref{hhwf})---that is, only the $q_G^n$. 
The boundary condition then gives rise to the equation
\beq
\sum_G\sum_{n=1}^2 {\bf H}_{\scriptscriptstyle(L,M)(G,n)}B_G^n=0
\eeq
where
\beq
{\bf H}_{\scriptscriptstyle(L,M)(G,n)}(k,\epsilon)=
\left\{ { {\bf m}_{{\scriptscriptstyle(L,M)G}}(k)g_1^{G,n}(\epsilon,R)-  
{\bf m}_{{\scriptscriptstyle(L,M)-G-1}}(k)v_1^{G,n}(\epsilon,R)\atop 
{\bf m}_{{\scriptscriptstyle(L,M)G}}(k)f_1^{G,n}(\epsilon,R)-
{\bf m}_{{\scriptscriptstyle(L,M)-G-1}}(k)u_1^{G,n}(\epsilon,R)}
\right\} 
\hbox{for} \left\{ \hbox{$L$ even} \atop   \hbox{$L$ odd}\right.
\eeq
and $\bf m$ is defined in Eq.~(\ref{matrix}).

\def\bkf{\hfill\break}

\end{document}